# Who Saves us From Risk? Altruists Promote Cooperation in a Public Investment Game

Shen Zhang[1,2,3], Xueyi Shen[1,2,3], Ruida Zhu[1,2,3], Zilu Liang[1,2,3] & Chao Liu[1,2,3*]

[1]State Key Laboratory of Cognitive Neuroscience and Learning & IDG/McGovern Institute for Brain Research, Beijing Normal University, Beijing 100875, China

[2]Beijing Key Laboratory of Brain Imaging and Connectomics, Beijing Normal University, 100875, Beijing, China

[3]Center for Collaboration and Innovation in Brain and Learning Sciences, Beijing Normal University, 100875 Beijing, China

*Correspondence concerning this article should be addressed to:

Chao Liu, Ph.D. E-mail: liuchao@bnu.edu.cn
State Key Laboratory of Cognitive Neuroscience and Learning & IDG/McGovern Institute for Brain Research
Beijing Normal University
19 Xinjiekou Wai Street, Beijing 100875, P. R. CHINA

**Abstract**

Providing commons in the risky world is crucial for human survival, however, suffers more from the "free-riding" problem. Here, we proposed a solution that limits the access of the resource to an agent and tested its efficiency with a novel public investment game. On each trial, all group members invest in an agent responsible for a gamble and distribution. Resources are distributed evenly between non-agent group members after the agent extracts for him/herself. In 3 laboratory experiments (n = 704), we found that, as an agent, many participants extracted fewer resources than others. This altruistic distribution strategy improved cooperation levels. Furthermore, this cooperation-promoting effect could be spread to selfish agents who are in the same group as the altruistic agent and was replicated in a one-shot setting. We proposed that, when the commons are under risk, this distribution solution serves as a better alternative to the well-documented sanction solutions.

**Introduction**

Providing public goods through cooperation is prevalent in human beings. As early as about 10,000 years ago, humans lived in small, mobile bands and made a living by hunting and gathering. Food was brought back to a central location and shared widely with other band members, most of whom were genetically unrelated [1,2]. In the modern world, cooperation takes a variety of forms. Public infrastructures such as dams, water supply, and street lamps, which are provided by all members of the society, are vital for our lives. Though crucial for human beings, public goods suffer from the "free riding problem" due to their non-excludability. That is, a rational individual would take advantage of others by not providing while still consuming the commons, which leads to inevitable impairment of public goods [3–5].

Moreover, one crucial task for humans is to cope with the risk that lies in the nature of the environment. Due to the risk and uncertainty in the environment, the returns of cooperation are often unpredictable. In hunting-gathering societies, the amount of meat gained by hunting is more variable and thus always not enough for our survival, to offset this risk, gathering serves as a complement to provide enough food [1]. Even in the modern world, natural risks such as floods or hurricanes could destroy human infrastructure, sometimes leading to a great cost of commons. Risk's detrimental effect on the commons may undermine the motivation to provide the commons, and indeed, the risk environment has been reported to significantly lower the cooperation level in groups [6–9]. However, despite its fatal detrimental effect on cooperation, few studies have focused on how to promote cooperation when commons are at risk.

Previous studies have identified sanction systems as an effective solution to the free riding problem. Aiming to decrease the fitness of the free rider, the sanctioning system can take the form of costly rewards or punishments by group members [10–12], partner exclusion [5,13–15], or resource allocation [16,17]. However, although the sanction system is efficient in promoting cooperation when it is present, two caveats may impair its efficiency in promoting cooperation under risk. First, the trust between group members has already been destroyed under risk conditions, while the sanction system may further lower the trust level as shown in previous literature[18,19]. Since trust was identified as a key factor influencing cooperation [20], the sanction system may not have enough power to restore the decrease in cooperation under risk. The second disadvantage of the sanction system is that it is always costly, both to the individual and to the group. As commons are already damaged under risk, punishment may thus be rare and weak.

In this paper, we instead adapted a distribution solution to promote cooperation under risk. In social interactions, when individuals are responsible for the harm of others, compensation would be a route to repair the relationship between the transgressor and the victim. When compensation is not possible, the transgressors even tend to punish themselves due to the guilt for accidentally harming others or the motivation of reputation repair [21]. We leveraged this idea in a public goods setting and hypothesized that when responsible for the loss of commons, individuals would sacrifice themselves to compensate the groups. Specifically, a group member would be responsible for the distribution of the resources. We further altered the structure of the distribution so that resources can only be distributed equally between members that cannot access the resource. This setup removes the possibility of punishing or rewarding specific individuals. Furthermore, we ensured that the contribution of the group members was private information that could only be accessed by the players themselves so that the reputation related to contribution is minimized.

These ideas were formalized by the public investment game (PIG), which is built on previous work with public goods games and dictator games, while introducing a gambling mechanism to model the risk in the environment to the commons. Public goods games have been widely used to model voluntary cooperation, and dictator games have been used as a tool to probe pure altruism. In the PIG, four players form a group, and each player is endowed with 100 units in the beginning. On each trial, one player is assigned the role of the agent, who will be responsible for the gamble and distribution of the public resource. Next, all players can invest 1-20 units from their private account into the public pool. The agent then plays a card-choosing game with the resources in the public pool. Upon choosing a red card (win), the public resource would be tripled, while choosing a black card would halve the resource (lose). Last, the agent extracts the resources for him/herself from the public pool, and the rest is distributed equally among the other players (Figure 1A, see Supplemental Information for details).

The investment and distribution mechanisms introduced in the PIG enable us to probe the effect of the distribution strategy on the group cooperation level.

Note that the settings of win and loss in the PIG describe two profoundly different payoff structures. Considering that the resources are distributed equally among the players, on winning trials, the PIG is reduced to a classical public goods dilemma in which players are better off when all players contribute 1 unit than when all contribute nothing. However, on losing trials, even if the remaining resources are distributed equally, all players contribute 1 unit leads to a loss for all. Hence, to maintain the payout of non-agent players, the agent must compensate them at their own cost. In addition, introducing the gamble and distribution mechanism does not influence the game-theoretic predictions of the game. Contributing nothing by all players would always be the Nash equilibrium, which leads to nothing to distribute.

We explored the distribution pattern and its effect on cooperation in the PIG in a series of three experiments. Specifically, in experiment 1, we tried to explore whether people act in an altruistic distribution manner and whether this distribution strategy has a positive effect on promoting cooperation. Experiment 2 replicated the results of experiment 1 with a more controlled setting and tested whether the promoting effect was only evident on trials with altruistic agents. Experiment 3 was designed to test whether the altruistic distribution and its promotion effect on cooperation fully come from concern for reputation.

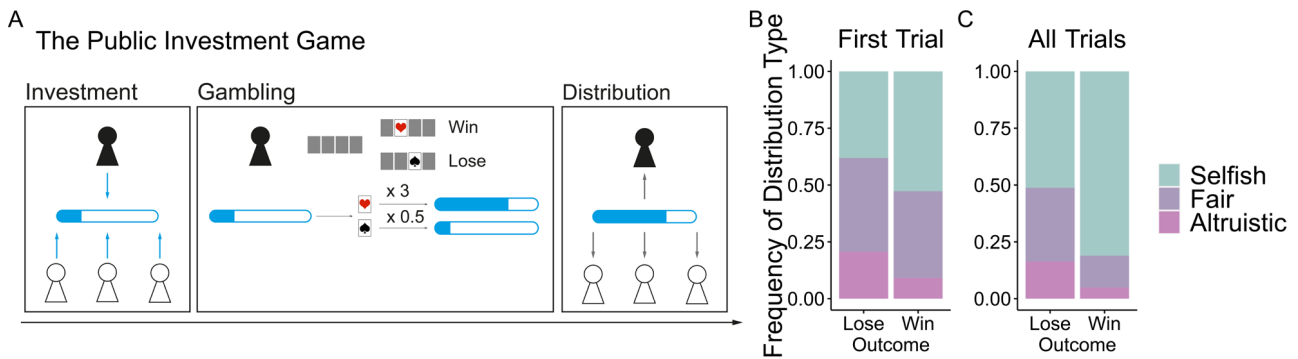

*Figure 1. The public investment game. A) The PIG consists of three phases. In the investment phase, an agent is chosen, and all players invest into the public pool. In the gambling phase, the agent plays a gamble in which a loss halves the resource in the public pool, while a win triples it. In the distribution phase, the agent distributes the resources in the public pool to all players. B) The frequency of the distribution strategies.*

# Experiment 1

Our first experiment aimed to explore the causal effect of distribution strategy on cooperation in a two-session interactive design with intervention. The same anonymous participants (denoted by the same letters during the task) played 2 sessions of the PIG with 40 trials each. The role of agent was assigned to each player in turn. The intervention was introduced after the first session.

Interestingly, we found that in the first session, on a considerable number of losing trials, the agent extracted the resource less than the group average (i.e., distributed altruistically; Figure 1B, on the first trial, Altruistic: 20.6%, Fair: 41.3%, Selfish: 38.1%; Figure 1C, overall, Altruistic: 16.4%, Fair: 32.4%, Selfish: 51.2%). On winning trials, almost all agents distributed the resource selfishly in the long run (Figure 1C, Altruistic: 5.0%, Fair: 14.0%, Selfish: 81.1%). Thus, in the following analysis, we only focused on the promoting effect of the distribution strategy on losing trials.

To explore the effect, groups were classified based on their members' dominating distribution strategy in the first session (see the Methods for details). Specifically, agents were classified into three types: altruist prefers to extract less than the group average on losing trials, individualist prefers the opposite, and egalitarian prefers to take exactly the same as others. As we predicted, for groups "with altruist" (n = 44), the investment level is significantly higher than that of groups of "four individualists" (n = 40) and that of "mixed" groups (n = 34, groups that are neither "with altruist" nor "four individualists") in the later round (one round is four trials in which each player acted as agent once) of session one (Table 1, Figure 1A).

Table 1 Investment level of non-agent players in session 1.

| Round | | R1 | R2 | R3 | R4 | R5 | R6 | R7 | R8 | R9 | R10 |
|---|---|---|---|---|---|---|---|---|---|---|---|
| With Altruist | | 8.07 | 7.84 | 7.65 | 7.59 | 7.47 | 7.52 | 7.16 | 7.08 | 7.17 | 6.90 |
| | | (0.25) | (0.38) | (0.39) | (0.45) | (0.42) | (0.47) | (0.47) | (0.50) | (0.49) | (0.46) |
| Mixed | | 7.46 | 6.90 | 6.17 | 5.40 | 5.27 | 4.96 | 5.02 | 4.90 | 4.32 | 4.37 |
| | | (0.33) | (0.33) | (0.40) | (0.38) | (0.41) | (0.45) | (0.46) | (0.50) | (0.49) | (0.47) |
| Four Individualists | | 7.31 | 6.17 | 5.23 | 4.86 | 4.45 | 4.19 | 3.70 | 3.61 | 3.37 | 3.43 |
| | | (0.31) | (0.38) | (0.39) | (0.36) | (0.35) | (0.35) | (0.30) | (0.29) | (0.29) | (0.32) |
| With an Altruist vs. Mixed | t | 1.44 | 1.87 | 2.66 | 3.76 | 3.74 | 3.94 | 3.26 | 3.08 | 4.14 | 3.86 |
| | p | 0.15 | 0.066 | 0.010 | <0.001 | <0.001 | <0.001 | 0.002 | 0.003 | <0.001 | <0.001 |
| With an Altruist vs. | t | 1.89 | 3.14 | 4.37 | 4.76 | 5.51 | 5.67 | 6.25 | 6.04 | 6.71 | 6.14 |

| Four Individualists | p | 0.063 | 0.002 | <0.001 | <0.001 | <0.001 | <0.001 | <0.001 | <0.001 | <0.001 | <0.001 |

To causally explore this effect, we implemented an intervention after the first session. Specifically, a confederate was introduced in session 2 to take the place of one chosen participant (see the Methods). For the "with altruist" group, the most altruistic player in session 2 (in terms of distribution selfishness on losing trials) was replaced by an individualist confederate. For "mixed" groups, a randomly chosen participant was replaced by an individualist confederate. For "four individualists" groups, the most selfish participant was randomly replaced by an altruist or an individualist confederate. Overall, confederates simulated different distribution strategies on losing trials, while keeping other behaviors the same. This manipulation enables us to further test the effect of distribution on the group investment level, controlling for confounders such as the distributions on winning trials or investments. To our surprise, we found no effect of confederates' distribution strategies on the investment level among the real participants (i.e., on trials when the confederate was not the agent). In the "four individualists" groups, there was no difference in group investment between the altruist confederate group and the individualist confederate group in all rounds (Figure 2B; all ps > 0.60). In fact, comparing the first and last rounds, the investment level of the participants does not change at all in all types of groups (Figure 2B, all ps > 0.17).

Cooperation level in Session 2 among participants suggests that when an environment of cooperation was settled, the cooperation levels between original group members thus persisted and were not influenced by the newly introduced confederates. We thus hypothesized that participants can actually discriminate different confederates who are newcomers. To test this, we considered the participants' investment when the confederate was the agent, that is, participants' investment in the confederate. Indeed, for the four individualist groups, the altruist confederate maintained participants' cooperation level (i.e., received higher investment) compared to the individualist confederate (Figure 2C and Table 2). Comparing the first and last rounds, the individualist confederate decreased the participants' cooperation level both in groups "with altruist" (7.47±0.52 (Round 1) vs. 5.14±0.54 (Round 10), t(43) = 5.74, p < 0.001) and "mixed" groups (5.46±0.56 (Round 1) vs. 4.11±0.60 (Round 10), t(33) = 2.65, p = 0.012).

We next tested whether the confederates' distribution strategy affected those of the participants. To do this, we compared the averaged distribution selfishness for all participants in the first and second half of session 2 (as a reminder, we focused only on the selfishness of losing trials. For illustration, Figure 2D plots the distribution selfishness summarized across groups that lost the gamble on some trials in that round). For the four individualist groups, an altruist confederate decreased distribution selfishness (2.48±0.45 (first half) vs. 1.74±0.42 (second half), $t(20) = -3.85$, $p < 0.001$). An individualist did not change the group-level selfishness for all group types (all $ps > 0.53$). These results indicate that for a stable group, only one individualist is not enough to change the group behavior.

**Table 2.** Investment in different confederates in the "four individualists" groups.

| Confederate Type | | R1 | R2 | R3 | R4 | R5 | R6 | R7 | R8 | R9 | R10 |
|---|---|---|---|---|---|---|---|---|---|---|---|
| Altruist | | 4.51 (0.75) | 4.90 (0.81) | 4.71 (0.73) | 4.86 (0.76) | 5.02 (0.72) | 4.89 (0.71) | 4.81 (0.73) | 4.73 (0.73) | 4.84 (0.80) | 5.03 (0.81) |
| Individualist | | 3.98 (0.44) | 4.00 (0.39) | 3.59 (0.37) | 3.56 (0.44) | 3.23 (0.39) | 3.33 (0.38) | 3.12 (0.39) | 2.98 (0.38) | 3.63 (0.45) | 3.05 (0.40) |
| Altruist vs. Individualist | t | 0.60 | 0.98 | 1.36 | 1.47 | 2.18 | 1.94 | 2.05 | 2.097 | 1.32 | 2.20 |
| | p | 0.55 | 0.33 | 0.18 | 0.15 | 0.037 | 0.06 | 0.049 | 0.045 | 0.19 | 0.036 |

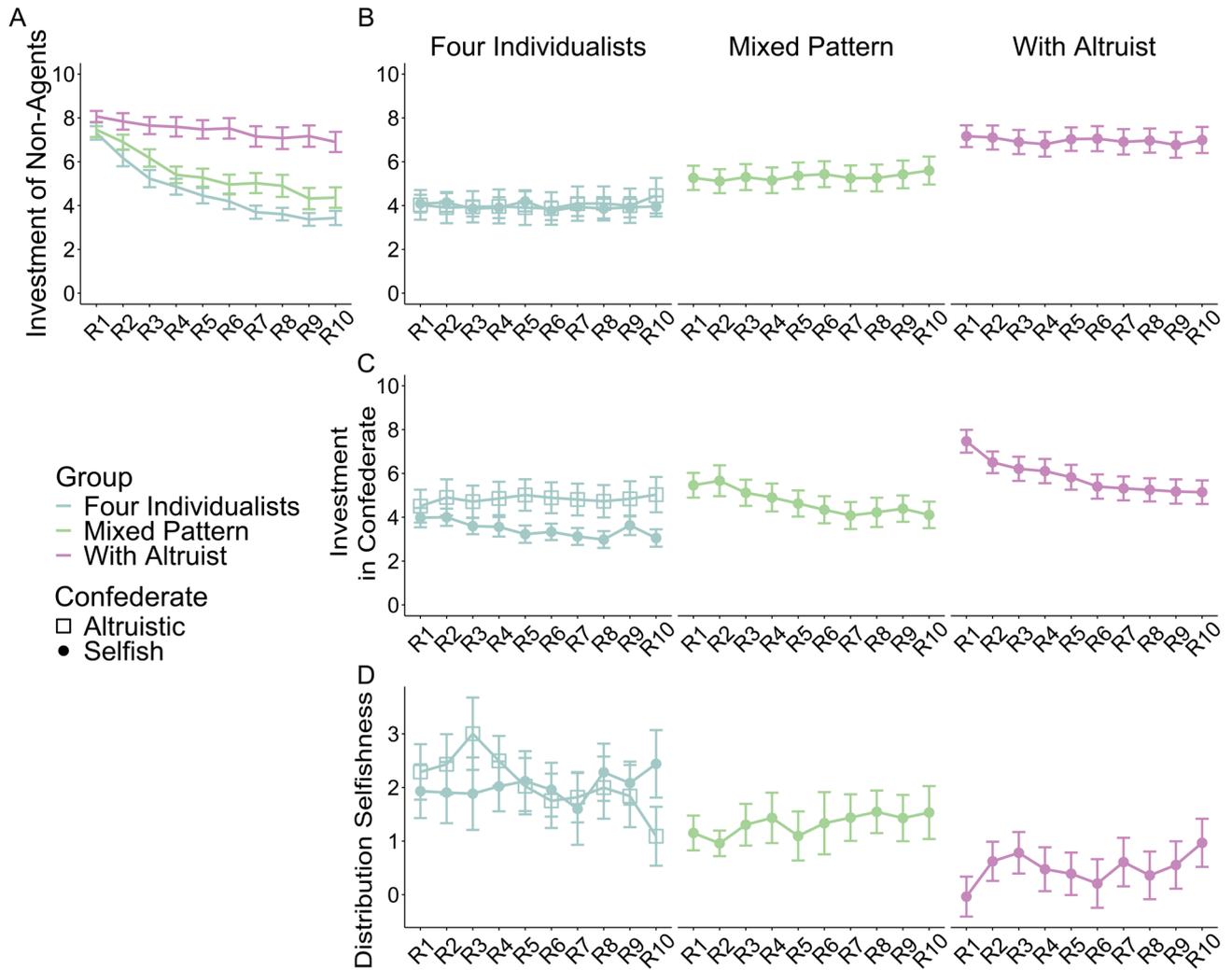

*Figure 2. Investment of each round in the public investment game. A) Non-agents' investment in Session 1; a higher level of investment was found in the groups with an altruistic agent. In session 2, B) all types of groups maintained their investment level. C) Investment in the confederate decreased in "with altruist" groups and "mixed" groups where an individualist confederate was introduced. For the "four individualists" groups, investment in the altruist confederates was higher than that in the individualist confederates. D) Distribution selfishness changed only when an altruistic confederate was introduced. All error bars indicate s.e.m.*

**Experiment 2**

We showed that in experiment 1, the altruist confederate helps to maintain the investment level. However, our groups in experiment 1 were randomly formed, and it is possible that the promotion effect of altruist agents we found in the altruist group was due to all of them being altruists at the beginning. Thus, in experiment 2, we want to show that in a more controlled design, the investment promotion effect elicited by altruistic agents can be found in preset groups and is not limited to the trials with altruist agent per se. To do this, we simulated data of a 40-trial PIG mimicking two social environments based on experiment 1. Participants were randomly assigned to either the one-altruist environment where one of the confederates was altruist or the three-individualist environment where all confederates were individualists (see the Methods). We expected that in the one-altruist environment, the investment level would be higher than that in the three-individualist environment.

Consistent with our expectations, we observed significantly higher investment level in the one-altruist than three-individualist environment in later rounds (Figure 3A, Table 3). We next aimed to test whether the effect was specific to altruists. To do this, we compared the investment level on trials of altruist agents against those of individualist agents in the one-altruist environment in groups with at least an altruist and individualists (Figure 3B, Table 4). Interestingly, participants invested more in altruist agents than in individualist agents in the middle rounds. In later rounds, the difference disappeared (Table 4) as investment in individualist agents was maintained.

Similar to experiment 1, we next explored the effect of the environment on distribution selfishness (Figure 3C). Again, the distribution selfishness showed marginal decrease in the one-altruist environment (0.73±0.41 (second half) vs. 1.16±0.41 (first half); $t(30) = -1.81$, $p = 0.081$) and marginal increase in the three-individualist environment (3.35±0.40 (second half) vs. 2.40±0.40 (first half); $t(29) = 1.94$, $p = 0.063$), resulting in a significant interaction ($F(1,59) = 6.90$, $p = 0.012$).

*Table 3.* Investment level in two social environments.

| Environment | | R1 | R2 | R3 | R4 | R5 | R6 | R7 | R8 | R9 | R10 |
|---|---|---|---|---|---|---|---|---|---|---|---|
| One Altruist | | 7.37 (0.55) | 7.20 (0.70) | 6.18 (0.65) | 6.19 (0.64) | 6.02 (0.73) | 5.75 (0.72) | 5.74 (0.73) | 5.51 (0.73) | 5.69 (0.76) | 5.25 (0.76) |
| Three Individualists | | 6.85 (0.56) | 6.56 (0.51) | 5.71 (0.62) | 5.23 (0.80) | 5.03 (0.70) | 4.74 (0.70) | 4.57 (0.73) | 3.65 (0.55) | 3.64 (0.56) | 3.46 (0.52) |
| Altruist vs. Individualist | t | 0.65 | 0.74 | 0.53 | 0.94 | 0.98 | 1.02 | 1.13 | 2.04 | 2.17 | 1.95 |
| | p | 0.26 | 0.23 | 0.30 | 0.17 | 0.17 | 0.16 | 0.13 | 0.023 | 0.017 | 0.0282 |

*Table 4.* Investment in different confederates in one-altruist environment.

| Confederate | | R1 | R2 | R3 | R4 | R5 | R6 | R7 | R8 | R9 | R10 |
|---|---|---|---|---|---|---|---|---|---|---|---|
| To Altruist | | 6.79 (0.54) | 6.76 (0.78) | 6.10 (0.79) | 5.79 (0.81) | 6.17 (0.85) | 6.17 (0.85) | 6.07 (0.99) | 5.38 (0.78) | 5.24 (0.85) | 4.82 (0.81) |
| To Individualists | | 7.03 (0.63) | 7.16 (0.72) | 6.22 (0.71) | 5.21 (0.61) | 5.56 (0.72) | 5.03 (0.76) | 5.12 (0.80) | 5.00 (0.79) | 5.28 (0.87) | 4.52 (0.80) |
| Altruist vs. Individualists | t | -0.50 | -0.92 | -0.29 | 1.14 | 1.7 | 2.07 | 1.09 | 0.64 | -0.045 | 0.53 |
| | p | 0.689 | 0.818 | 0.614 | 0.132 | 0.050 | 0.026 | 0.143 | 0.263 | 0.518 | 0.301 |

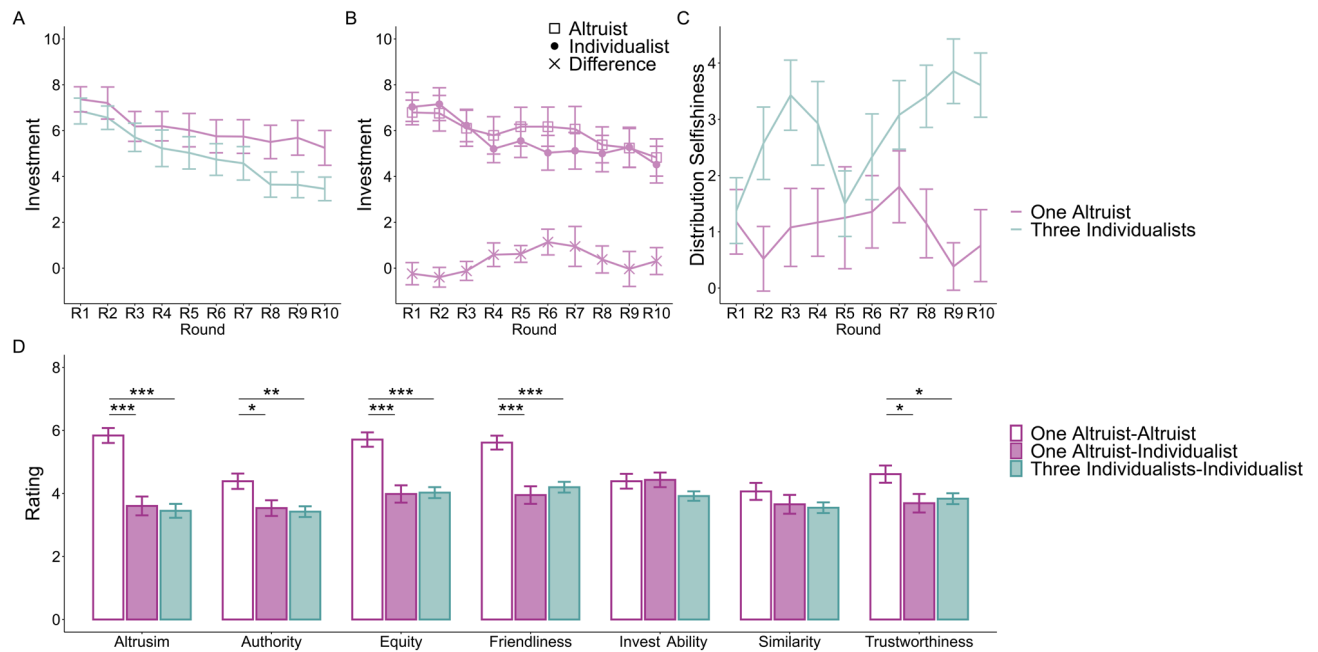

*Figure 3. Generalizing the effect of altruist. A) Investment was higher in one-altruist environment than in the three-individualist group. B) Investment in the individualist in the one-altruist environment was lower in the middle of the session and was maintained in later rounds. C) Distribution selfishness was moderated by social environments, where selfishness in three-individualist increased and selfishness in the one-altruist group decreased. D) Participants' ratings on confederates. The altruists gained a better reputation than the individualists. * p < 0.05, ** p < 0.01, *** p < 0.001. All error bars show s.e.m.*

Moreover, we asked the participants to rate all confederates on a series of trait-related questions. The results showed that the altruist confederates gained a better reputation than the individualist confederates (Figure 3D), indicating that participants kept tracking the behaviors of each confederate instead of mixing them up.

**Experiment 3**

In experiment 2, we showed that the promotion effect of the altruist on investment is not limited to trials when the altruist is an agent. All these findings can be attributed to the interactive nature of the first two studies. Thus, in experiment 3, when no reputation was involved, we wanted to explore whether participants still adopted the altruistic distribution strategy and whether it still boosted cooperation. To this end, we leveraged a one-shot design in which participants were asked to play three trials of PIGs. The participants were told that they would encounter each player only once. We manipulated all trials to be losing trials. The participants were agents on the first and third trials, while on the second trial, another random confederate was the agent and distributed the resources altruistically (see the Methods). We hypothesized that after exposure to the altruistic distribution, the investment level of participants would increase on the third trial.

A total of 111 participants were recruited in experiment 3, and we replicated our findings in experiment 1. Twenty-seven (24.3%) participants distributed the resource altruistically after causing the first loss, while 46 (41.4%) participants chose to allocate equally across group members, leaving the remaining 38 (34.2%) participants who acted selfishly.

Consistent with our hypothesis, the investment level on the third trial was significantly higher than that on the first trial (Figure 4A, 8.16±1.13 on the first trial vs. 9.16±1.60 on the third trial, t(110) = -3.89, p<0.001). Such an increase was more salient in the equal (7.78±1.22 on the first trial vs. 9.17±2.71 on the third trial, t(45) = -2.90, p=0.006) and selfish groups (8.82±2.51 on the first trial vs. 9.61±2.78 on the third trial, t(37) = -2.45, p=0.019) than in the altruistic group (7.89±2.56 on the first trial vs. 8.52±2.78 on the third trial, t(45) = -1.27, p=0.22, Figure 4B).

However, it is possible that participants increased their contribution because they became aware of their dominant right to allocate themselves more when they were agents and therefore invested more. To exclude this explanation, we compared the distribution on the first and third trials and found no difference for all distribution strategies (Figure 4C; D, all ps > 0.50).

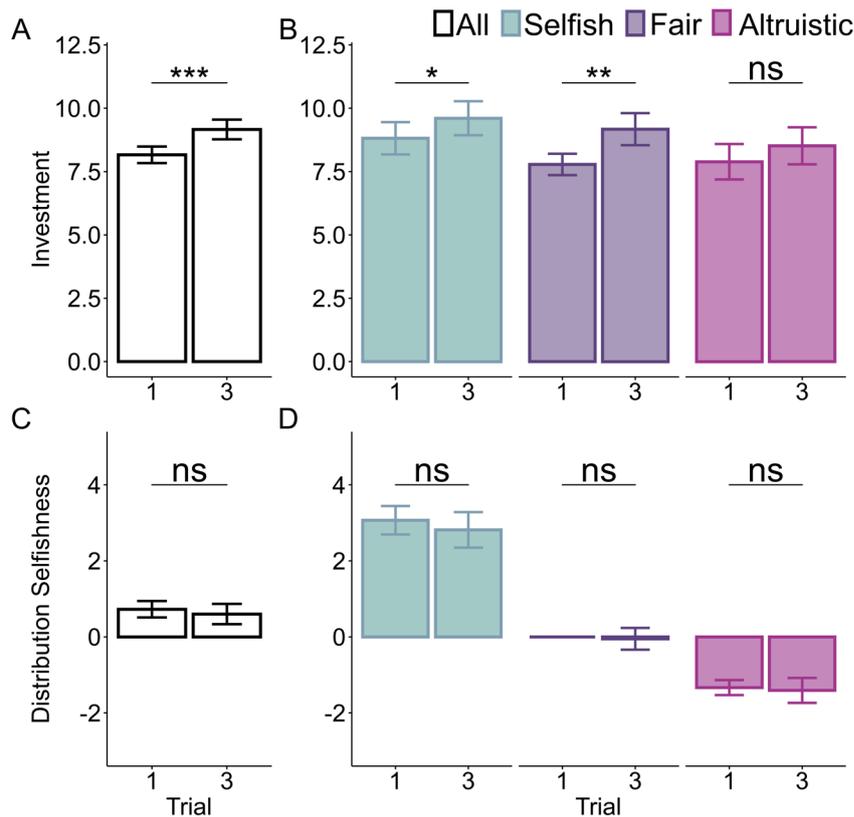

*Figure 4. Altruist effect when reputation concern was excluded. a) Investment levels for all participants. The investment increased from the first to the third trials. b) Investment conditioned on participants' first trial distribution. c) The distribution strategy did not change from the first to the third trial. d) Investment conditioned on the first trial distribution strategies. Errorbars indicate s.e.m.*

## Discussion

Through three experiments, we showed that in a novel public investment game, when responsible for the impairment of public resources, a considerable number of participants distributed the resources altruistically. Compared to the selfish distribution strategy, distributing altruistically significantly increases the cooperation level. This promotion effect was not limited to the altruist agent themselves and could be found not only in random groups (experiment 1) but also in preset groups (experiment 2). In addition, the altruistic distribution was found in a one-shot design where the possibility of repeated encounters was totally removed (experiment 3).

Several studies have shown that when leaders are responsible for allocating resources, they usually take more than other members [22–26]. Though on average, participants in all three experiments distributed the resources selfishly both when winning or losing the gamble, we found a considerable number of participants took fewer than other members on losing trials (Figure 1C). The reason why the participants adopted the altruistic distribution strategy may be mixed.

The image score is a main concern in indirect reciprocity [27], and people try to maintain their reputation by providing commons [28–31]. When responsible for the risk in the environment, participants try to save his/her reputation from losing money by acting altruistically. However, in experiment 3, when the players could meet only once, we still found a considerable number of participants who distributed the resource altruistically. Thus, the altruistic strategy can only be partially explained by indirect-reciprocity accounts.

The social heuristic hypothesis [32–34] proposed that individuals used intuition they acquired from previous social interactions as an approximation of deliberate calculation for the current environment. We showed in experiments 1 and 2 that the altruist elicited a significant increase in investment not only to the group member him/herself but also, to some extent, to the group he/she was in. These positive interactions may shape the intuitional response when deciding how to distribute the resource. Thus, although participants did not have the opportunity to repair their image, the intuitional response to the loss would always be to compensate others even at the cost of their own.

Another explanation related to intuitional account (i.e., the SHH) is the spontaneous emotional response account. A large body of researchers have proposed that the irrational rejection of unfair offers in ultimatum games comes from negative emotions such as anger or envy [35–37]. More recently, research on interpersonal emotion showed that when accidentally harming others, the transgressor would even punish themselves to show remorse or guilt [21,38] and try to compensate the victims [39,40]. As in our settings, the agents are responsible for the loss of the gamble, and this unintentional harm to others may induce these interpersonal emotions and may thus cause the seemingly irrational altruistic distributions in our experiment 3.

The last potential explanation of the altruistic distribution is social conformity [41]. That individuals follow others' behavior has long been documented [42–45]. It was reported recently that even a single model would be enough to change the observers' behavior [46–48]. The distribution strategy changes in experiment 1 and experiment 2 are consistent with this explanation. However, in experiment 3, we did not find any changes in distribution selfishness after one trial of exposure to an altruist agent. This may be because the change in distribution may thus need enough interaction to take effect. Note that these factors are not mutually exclusive, and future studies are needed to dissociate their contributions and delineate the boundaries of when these factors are involved.

One of the most considered factors influencing cooperation is the payoff. Several institutional solutions have been proposed to improve cooperation by changing payoffs to reward cooperators and/or to punish noncooperators. This sanction system can take the form of direct punishment and reward from group members [11,20,49,50], negative or positive reciprocity in another game [28,29,31], or resource distribution by a leader [22–25,51]. All these solutions require direct reputation information of the transgressor to take effect. Rational individuals who try to maximize payout would choose to cooperate if there is a risk of punishment or to try to gain the reward. In our study, the payout maximization account can be expressed as gain by distribution; that is, on trials with altruistic agents, payout of other group members is somewhat guaranteed under altruistic distribution. Our finding of a higher level of investment level in an altruistic agent (compared to a selfish agent) in session 2 of experiment 1 is consistent with this account. However, we found in experiment 3 that participants' investment was higher after exposure to an altruist agent and that they were not concerned about the payout, as their distribution strategy was quite stable. Furthermore, we found that in experiment 2, the investment increase is not only limit to the altruistic agent and in experiment 3, even when repeated interactions are totally excluded. Thus, the gain by distribution account cannot fully explain the promotion effect of cooperation.

The second explanation is trust. The altruistic distribution of the resource is a costly signal of a positive attribute, indicating a cooperative tendency. Thus, the cooperation level would be increased through direct reciprocity. This argument is supported by the attitude ratings towards different agents in experiment 2. Many studies have demonstrated that individuals who trust other members in the group contribute more [19,49,52,53]. The trust of a person may be generalized to people in general. Thus, the cooperation level increase in experiments 2 and 3 is consistent with this account.

As mentioned before, the institutional solutions of the "free riding" problem always come in the form of sanction systems that reward cooperators and punish defectors. However, sanction systems suffer from the problems of antisocial punishment, where noncooperators punish cooperators even at a cost [50,54–56]. As altruistic distribution takes the form of only the cost of the agents themselves, no revenge would be possible. Indeed, sanction systems were found to undermine interpersonal trust [18,19]. As mentioned before, trust is maintained in experiment 2.

Another disadvantage of sanction systems is that sanctions always form second-order public goods in which second-order free riders can reap the fruits of others' sanctions [30,56–58]. In experiment 2, the investment to the selfish agent was maintained in the one-altruist group, which may indicate that the individualist sometimes can also free ride on the reputation of the altruist. This finding suggests that the distribution solution may also suffer from the second-order free riding problem. However, to what extent the problem undermines public goods has yet to be tested.

The last flaw of sanction solution is that when the system is removed, the cooperation level decreases significantly [28,50,59]. In experiment 1, the investment level between participants was maintained even when an individualist confederate was introduced in session 2. This result may indicate that while an atmosphere of trust was formed through enough positive interactions, the cooperation level between original group members would thus persist. However, in experiment 1, the first session lasted for 40 trials, and it is still unknown how long a positive interaction is needed for trust between members to be formed.

In summary, in three experiments, we showed that when commons are at risk, a considerable number of participants tend to distribute the resource altruistically. This distribution tendency cannot be fully attributed to concern for reputation. The distribution strategy significantly influences the group cooperation level, possibly through the costly signal of trust, which can be generalized to all. The responsibility taking solution is exempt from revenge, which is present in most sanction solutions.

# Methods

## Experiment 1

### Participants

Five hundred and twenty-eight participants at Beijing Normal University completed the experiment (163 males, 365 females; 22.4 ± 2.5 years) and were randomly assigned to 4-member groups. Group members were previously unknown to each other, and interaction before the experiment was refrained by allocating participants to 4 separate rooms. Data from 5 groups were excluded due to failed avoidance of interaction before experiment, and another 9 groups were excluded because of experimental operation error; thus, 472 participants' data from 118 groups were included in the analysis. Monetary reward was given according to performance in the experiment. Protocols in all three studies were approved by the Institutional Review Board at the Beijing Normal University.

### Procedure

Four participants came to the laboratory at the same time. On arriving, each one was led to a small testing room separately to preclude any verbal interactions among them. After being revealed the nature of the public investment game, all subjects answered some check questions to confirm that they understood the task correctly. After one session of the public investment task, the participant replacement procedure (see the participant replacement for details) was implemented. Then, another session was run.

### Labeling and player replacement procedure

To explore the effect of distribution on investment, we classified groups according to their members' tendency of distribution. Specifically, the distribution strategy of the agent on a trial was labeled selfish, fair, or altruistic if the agent took more than, the same as, or less than other players on that trial. As the agents distributed the resource selfishly on almost all winning trials, we classified the players based only on their distribution on losing trials. That is, if the player adopted the same distribution strategy on more than 50% of losing trials when he/her played as an agent, then the player was labeled according to this dominating strategy (altruist (egalitarian, or individualist) for dominating altruistic (equal or selfish) strategy,). When no dominating strategy was identified, the player was labeled "not dominating". We then classified groups according to the dominating strategy of their members. Specifically, if there was at least one altruist member, then this group would be labeled as the group "with altruist". Otherwise, it would be labeled a group without an altruist. The group without altruistic players was further classified as "four individualists" if all dominating distribution strategies of its members were selfish. Otherwise, it would be labeled "mixed".

We implemented the player replacement procedure to manipulate the distribution strategy of one player in the group and to causally explore the effect of its member's distribution strategy on the group investment level. Specifically, for groups with an altruist, the most altruistic individual would be replaced by an individualist. For "four individualists" groups, the most selfish individual was replaced randomly by an individualist or an altruist. For mixed groups, we randomly removed one of the players and replaced him/her with an individualist. All confederates invested 11 points per trial during the task, and the individualist confederates distributed to themselves 3 units more than fair, while the altruist confederates distributed to themselves 3 points less than group average. We explained the nature of the deceit in the experiment to participant who was to be replaced, and the confederate was introduced to the task when taking this player's place in the second session.

**Investment**

**Session 1**

The investment was used as a measure of cooperation. As we found that players invested more in the public pool when they themselves were agents, the investment was then considered separately for agents and other players on each trial. As the participants took turns to act as the agent, we averaged the investment in each group every four trials (i.e., one round). On each round, investment levels of different types of groups were compared using independent samples t-tests.

**Session 2**

As confederates were introduced in the second session, we expected the investment in altruist confederate to be higher than that of the individualist. To do this, in four individualist groups, we used independent samples t-tests to compare investments into different confederates in each round. We also tested whether this contribution-boosting mechanism can be generalized to players other than confederates. That is, we compared the investment of participants in each other in each round again using a two-sample t-test. For groups with an altruist and mixed groups, we expected a decrease in investment in the confederate and among the participants. To test this, the investments in the first round and last round were compared with a paired t-test.

**Distribution**

Furthermore, in session 2, we considered whether the distribution behaviors of players are affected by confederates. Specifically, the distribution selfishness (i.e., the difference between the payout of the agent and other players) on loss trials was compared between the second and the first half of session 2 using a paired t-test.

**Experiment 2**

**Participants**

We recruited 65 participants from Beijing Normal University (age 22.4 ± 2.5 years, 40 female). All participants gave consent to take part in the experiments.

**Procedure**

Data from experiment 1 are used to generate player behaviors in experiment 2. Specifically, in a one-altruist environment, for each participant, we first choose a group with only one altruist and use the behavior of the altruist and two other randomly chosen players in the group as the behaviors of participants' opponents. For the three-individualist environment, behaviors of three randomly chosen players in one four-individualist group were used.

Participants came to the laboratory in groups of two to four and were randomly assigned to the one-altruist or three-individualist environment. The participants were also introduced to the nature of the PIG and told that they would be playing the PIG while interacting with three other participants. Instead, upon arrival, each subject was led to one of four cubes in which they could not see each other. After the task, the participants were asked to rate their attitude towards the players they interacted with and to answer some debrief questions about the deceptive nature of the task.

**Experiment 3**

**Participants**

We recruited 111 participants (age 22.1 ± 2.7 years, 75 female) from Beijing Normal University. All participants gave consent to participate in the experiment. Five of the participants showed suspicion in the debrief questions, but when removing the data of these five participants, all results were the same. Thus, we kept all participants in the main text.

**Procedure**

Participants came to the laboratory in groups of 8 to 12. When completing the task, the participants could see each other but were not allowed to speak to each other. They were told that they would be interacting with others in the testing room and another testing room and could only interact with other players once. After the task, they were asked to answer some debrief questions about the deception used in the experiment.

**Data availability**

The data sets collected for these experiments and data analysis code are available from the corresponding author upon reasonable request.

**Acknowledgment**

This work was supported by the Scientific and Technological Innovation(STI) 2030-Major Projects (2021ZD0200500), the National Natural Science Foundation of China (32441109, 32271092, 32130045), the Beijing Major Science and Technology Project under Contract No.Z241100001324005，and the Opening Project of the State Key Laboratory of General Artificial Intelligence (SKLAGI20240P06).

**Acknowledgements:**
This work was supported by the National Key Research and Development Program of China (2017YFC0803402), the National Natural Science Foundation of China (NSFC) (31871094), the Major Project of National Social Science Foundation (19ZDA363), and the Beijing Municipal Science and Technology Commission (Z151100003915122).

**Author contributions**
S.Z., X.S., R.Z., and C.L. designed the experiment. S.Z. and X.S. programmed the experiment. S.Z., X.Z., R.Z., collected the data. S.Z. and X.S. performed the data analyses. S.Z., X.S, Z. L., and C.L. wrote the manuscript.

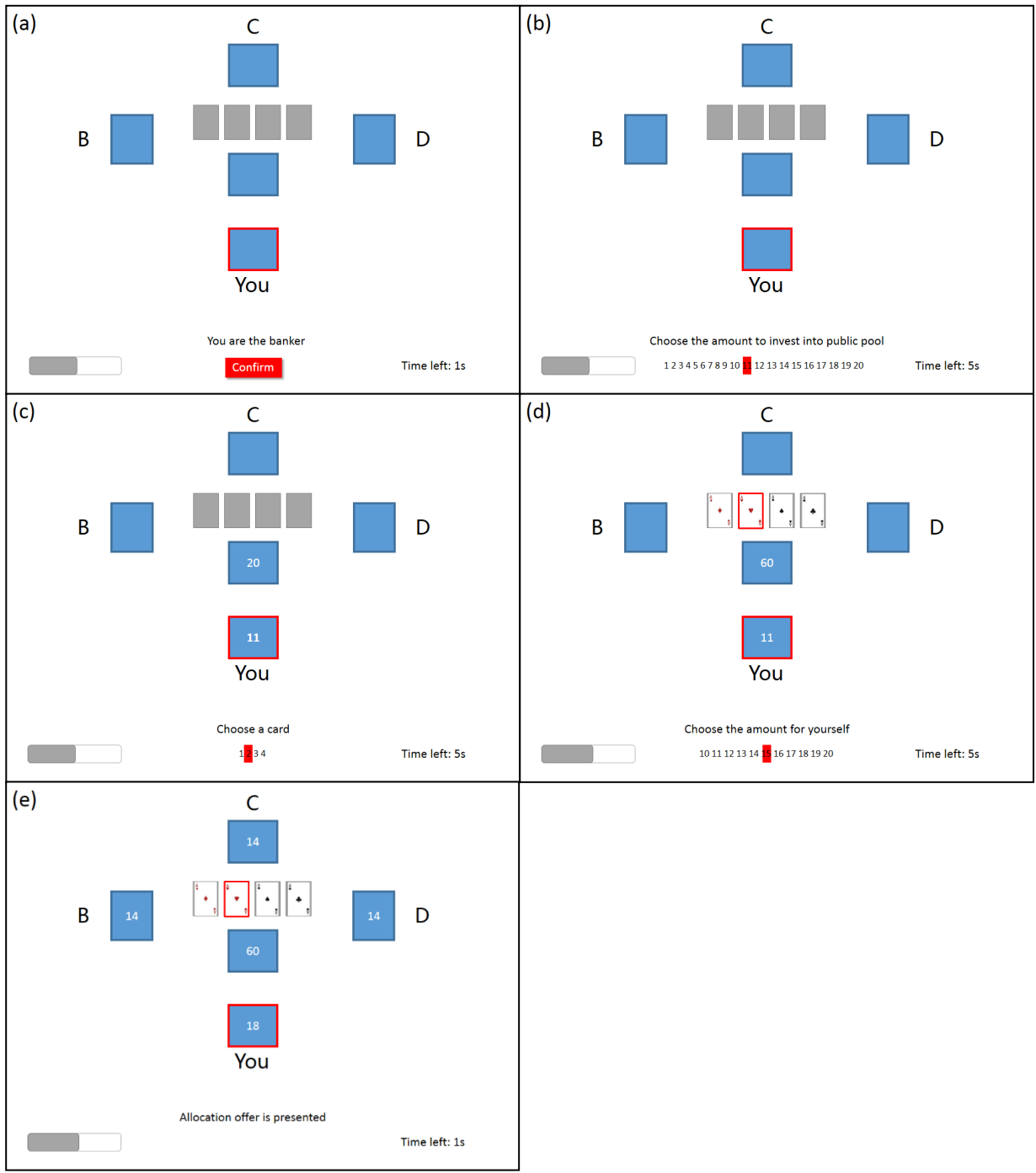

Figure S1 Public goods game procedure timeline of one trial. (a)Agent assignment, lasting for 1s. In this stage, one of the group members will be chosen as the agent. The deck of the agent would be highlighted by red rim throughout the trial. (b)Investment, lasting for 5s. Each player invests 1 to 20 unites from their individual accounts to the public pool. The cursor is initially on ten. Players choose the amount by pressing the left/right arrow keys on keyboard and confirm their choice by ENTER. If participants failed to press ENTER before timeout, the last choice which the cursor is on would be identified as their choice. At the end of the stage, players' own investment would be shown on their deck, while the amount of the public pool would be displayed on the public (central) deck. Investment of other players are not displayed. (c) Gambling stage, lasting for 5s. The agent should choose one card from four. If a red card was chosen, the group wins and the public resource rise by 200%, but if a black card was chosen, the groups loses and the public resource drop to 50%. For other players, they would be instructed to wait for the banker to make decision. (d) Distribution stage. Default option is to equally allocate among 4 players including the agent himself ($M_{Allocation}$), and other available options are integers ranging from ($M_{Allocation}$ -5) to ($M_{Allocation}$ +5). In case the banker loses and the lower threshold ($M_{Allocation}$ -5) reach below 0, the lower limit would be corrected into 0. (e) Outcome presenting session. Each player's allocated interest is presented on their individual deck.